# Synced Parallel Control Paths for Multi-Task System Operation

Behrooz Mirafzal, and Fariba Fateh
Kansas State University, Manhattan, KS, USA

## 1. Introduction

Numerous systems require the capability to switch their operational modes seamlessly without any disruptions. The proposed control architecture facilitates this transition smoothly, ensuring minimal transient effects. A prime example of this application is found in grid-interactive inverters, which need to alternate seamlessly between grid-following and grid-forming modes of operation [1]-[6], [32].

Smooth transitions with minimal transients are of paramount importance in systems like power subdivisions or critical buildings, especially when they form islanded microgrids after a blackout. This is crucial for maintaining the stability and reliability of the microgrid. When these subdivisions are isolated from the main utility grid, inverters play a vital role by smoothly switching from Grid-Following (GFL) to Grid-Forming (GFM) mode. This transition is essential to retain stable bus voltages and control the frequency of the microgrid. Rapid transitions can lead to power fluctuations, potentially causing equipment damage or operational failures. Moreover, managing static and dynamic loads efficiently is critical in such islanded systems. Inverters must work collaboratively to balance these loads, ensuring efficient energy distribution and preventing power outages. Smooth transitions are also imperative for maintaining frequency stability, particularly in low inertia systems where the frequency is more susceptible to fluctuations. The use of an adaptive controller facilitates this flexibility, allowing inverters to operate effectively in both modes, thereby enhancing the resilience of the power grid under isolated conditions. Furthermore, these smooth transitions help in reducing electrical stress on the equipment, prolonging its lifespan and reducing maintenance needs. Overall, ensuring minimal transients during mode switching is vital for islanded microgrids' operational integrity, reliability, and efficiency, preserving against instability and disruptions in supplying electricity.

## 2. Synced Parallel Control Paths: Concept and Design

*Synced Parallel Control Paths* as a concept can be applied broadly to various systems that require switching between different modes of operation, representing a substantial improvement over traditional control mechanisms. This architecture integrates two simultaneous sets of control paths – the primary and auxiliary paths – which regulate key operational parameters irrespective of the system's current mode. The primary paths are tasked with generating the essential operational references critical for maintaining the desired output characteristics of the system. In parallel, the auxiliary paths are designed to continually align their outputs with those of the primary paths, ensuring a harmonious and synchronized operation. This feature of continuous synchronization distinguishes it from conventional systems, where transitions between modes often lead to operational disruptions or transients. By reducing these transients, Synced Parallel Control Paths greatly enhance the stability and reliability of the system, an aspect that is particularly vital in applications where precision and continuity are paramount. This dual-path strategy not only facilitates smoother transitions during mode changes but also significantly boosts the overall robustness and adaptability of the system to varying conditions and requirements. This approach marks a notable advancement in control technology, applicable across a wide range of systems requiring dynamic mode switching.

This innovative control system features a dual-pathway design, enhancing its adaptability and efficiency. Unlike traditional systems, where primary and secondary roles are fixed, this system allows for interchangeable roles between its two paths. The main path is tasked with issuing critical operational commands, ensuring the system performs tasks with high accuracy and efficiency. In parallel, the secondary path continuously adjusts to stay in sync with the main path, promoting a seamless operational flow. This approach significantly differs from older control systems, which often struggle with delays or coordination issues during task transitions or in varied environments. The system improves stability and reliability by reducing these disruptions, especially in complex scenarios. The essential advantage of this system lies in its smooth handling of task transitions and rapid adaptation to changing conditions. This not only streamlines operations but also boosts the overall efficiency and adaptability of the system. This advancement is a notable step forward in control system design, emphasizing its suitability for dynamic and evolving environments.

The suggested approach is not confined to just two paths; however, a main path must always exist, while multiple auxiliary paths can exist and function in the background. The proposed control technique, referred to as *Synced Parallel Control Paths* for a dual-path system, is depicted in Fig. 1. For this system, the outputs of the control paths are formulated as follows:

$$X_{c1}(s) = \{w\,(R_1^* - R_1) + \bar{u}\,e(s)\}\,G_{c1}(s)$$

$$X_{c2}(s) = \{\bar{w}\,(R_2^* - R_2) + u\,e(s)\}\,G_{c2}(s)$$

where, $e(s)$, is the error between the parallel paths, and it can be expressed as follows:

$$e(s) = \left(u\,X_{c1}(s) + \bar{u}\,X_{c2}(s)\right)$$

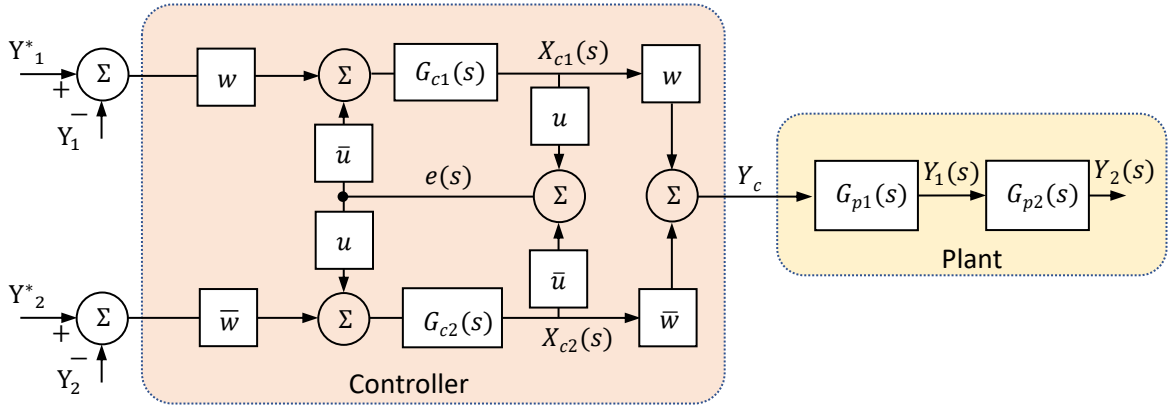

**Fig.1.** Graphical Demonstration of Synchronized Parallel Control Paths in Dual-Task System Operation.

The output of the controller, $Y_c(s)$, is written as:

$$Y_c(s) = w\,X_{c1}(s) + \bar{w}\,X_{c2}(s)$$

The error loop ensures that the output value of the background path remains consistent with the main path. This consistency is crucial at every switching moment, as it stabilizes the system and maintains seamless operation.

Typically, $G_{c1}(s)$ and $G_{c2}(s)$ are PI controllers. If they are not PI controllers, then there is a need to add an integrator in the error path, see Fig. 1.

The switching functions $(u, \bar{u})$ and $(w, \bar{w})$ are shown in Fig. 2.

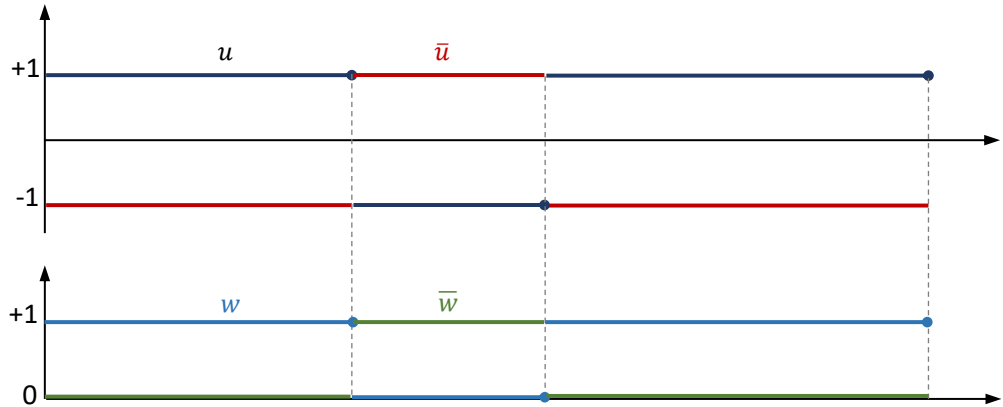

**Fig.2.** The Switching Functions $(u, \bar{u})$ and $(w, \bar{w})$ Used in the Controller.

**Example (1):** To illustrate the effectiveness of this controller, let us consider the plant transfer functions of $G_{p1}(s) = \frac{2}{s^2+4s+12}$ and $G_{p2}(s) = \frac{2}{s+4}$, and the desired outputs of $Y_1^* = 100$ and $Y_2^* = 50$. The PI controllers are also set to be $G_{c1}(s) = 2 + \frac{10}{s}$ and $G_{c2}(s) = 3 + \frac{18}{s}$. The switching functions have a period of $T = 50$ seconds and $T_w/T$ ratio is 0.7.

The selection of $G_{p2}(s)$ ensures that $Y_1$'s desired value of 100 corresponds to $Y_2$'s desired value of 50. Therefore, even when the controller alternates between the two paths, the output is expected to remain unaffected by this switching.

Fig. 3 shows the simulation results for both situations: one with the syncing loops and the other without them. To make the syncing loops ineffective, one can set $u = \bar{u} = 0$; however, the switching functions $(w, \bar{w})$ remain active in both scenarios.

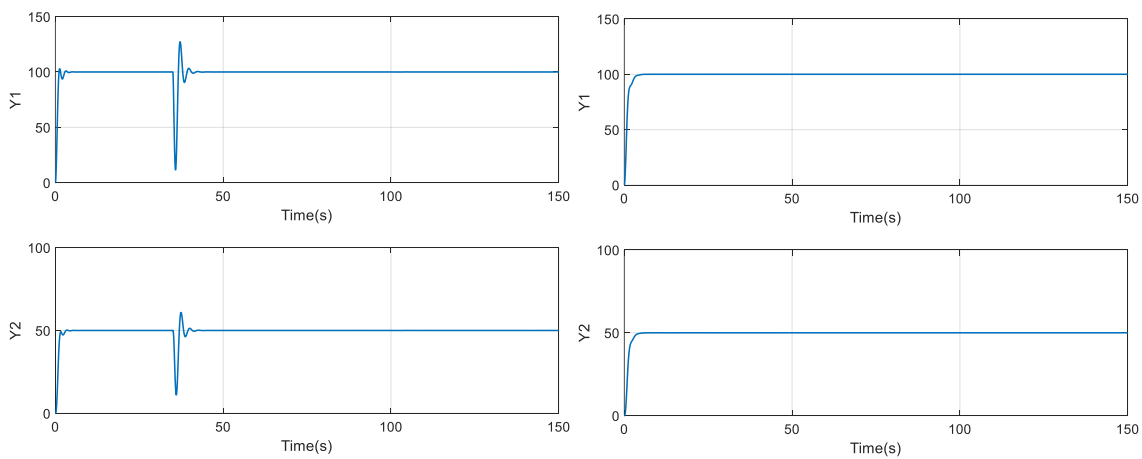

**Fig.3.** Comparison of Control System Performance: The right side of the image shows simulation results with synchronization loops, demonstrating efficient and stable system behavior. The left side illustrates results without synchronization loops, revealing irregular performance.

**Example (2):** To demonstrate the switching effects, let us only change $G_{p2}(s) = \frac{4}{s+4}$ and simulate for $G_{c2}(s) = 3 + \frac{18}{s}$ and $G_{c2}(s) = 3 + \frac{6}{s}$, when everything else is the same as given in Example (1). Fig. 4 shows the simulation results for both situations: one with the synchronization loops and the other without them.

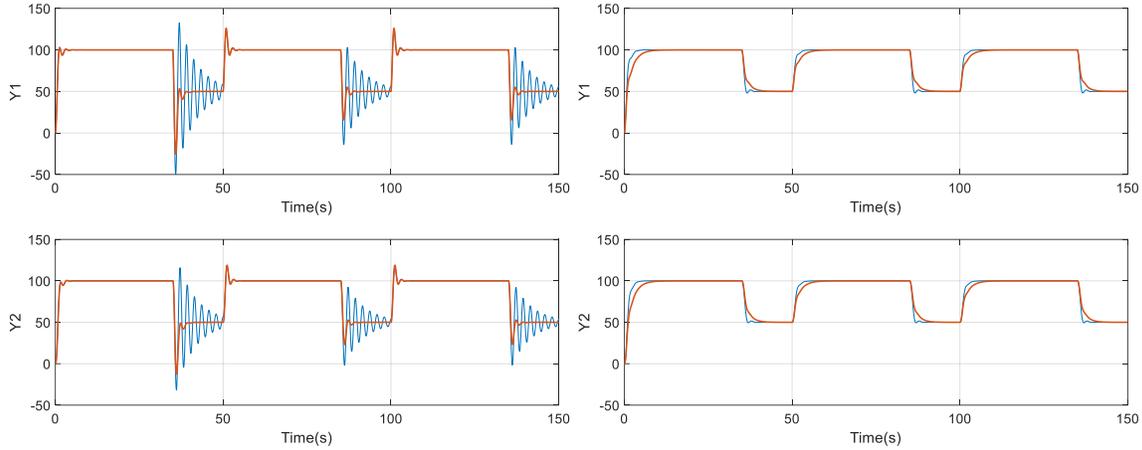

**Fig.4.** Simulation Outcomes Comparing Control Systems with and Without Synchronization Loops. On the right, results for a system using $G_{c2}(s) = 3 + 18/s$ (blue) and $G_{c2}(s) = 3 + 6/s$ (orange) with synchronization loops, showing coordinated and stable responses. On the left, the same system using again $G_{c2}(s) = 3 + 18/s$ (blue) and $G_{c2}(s) = 3 + 6/s$ (orange) without synchronization loops, depicting less stable and more oscillatory behavior.

## 3. Application of Synced Parallel Control Paths for Grid-Interactive Inverters

In modern power systems, the stability of inverters holds paramount importance due to several critical reasons [7]-[31]. As power grids increasingly incorporate renewable energy sources like solar and wind, inverters are essential for converting the variable direct current (DC) from these sources into a stable alternating current (AC) for grid compatibility. This conversion process needs to be efficient and reliable to maintain the quality and consistency of power supply. Moreover, inverters play a pivotal role in regulating voltage and frequency within the power system. Instabilities in inverters can lead to fluctuations in these parameters, potentially causing damage or malfunction of electrical equipment. Additionally, inverters are integral to the grid's overall support and reliability, offering crucial services such as voltage support, frequency control, and reactive power management [7], [24]-[29]. Their stability is vital in managing the intermittency of renewable sources, ensuring a steady electricity supply despite the inherent unpredictability of these energy sources. Inverters are also crucial in microgrids and island operation, where they facilitate smooth functioning, especially during transitions between grid-connected and isolated modes. Furthermore, they contribute to load balancing and improving power quality, addressing issues like harmonic distortion and voltage variations. With the dynamic and ever-changing nature of modern power grids, including the growing integration of energy storage systems, the stability of inverters becomes even more critical, ensuring the system's adaptability without compromising performance. Thus, inverter stability is a cornerstone in the effective and efficient operation of contemporary power systems, reflecting the evolving energy generation and consumption landscape.

This innovative method provides a sophisticated and dependable way to regulate power output and voltage waveforms, substantially improving conventional control systems [32]. This advanced design integrates two sets of parallel control paths – primary and auxiliary – that adeptly regulate voltage (V) and phase-angle (θ) in inverters, regardless of whether they operate in grid-following or grid-forming mode [1], [2], [3]. The detail of applying this control technique for grid-forming inverters has been reported in [1], see also Fig. 5. The primary paths are essential in generating precise references for the Pulse Width Modulation (PWM) generator, which is crucial for maintaining the inverter's desired electrical output. In parallel, the auxiliary paths ensure continuous synchronization with the primary paths, achieving a cohesive and smooth operation. This persistent synchronization starkly contrasts older inverter systems, where switching modes often lead to operational disturbances or transient issues. By curbing these disruptions, the Synced Parallel Control Paths significantly enhance the stability and reliability of the inverter's power supply, especially vital in environments where consistency is critical. Furthermore, this dual-path strategy not only facilitates smoother transitions between different operational states but also augments the overall efficiency and adaptability of the inverter, adapting seamlessly to changes in power system dynamics. This innovation marks a significant stride in enhancing inverter control technology.

Block Diagrams Illustrating the Proposed Control Scheme Applied for Inverters: The left diagram shows the phase-angle reference for inverters. The right diagram displays the voltage-amplitude reference, highlighting the versatility and adaptability of the control strategy in different operational modes.

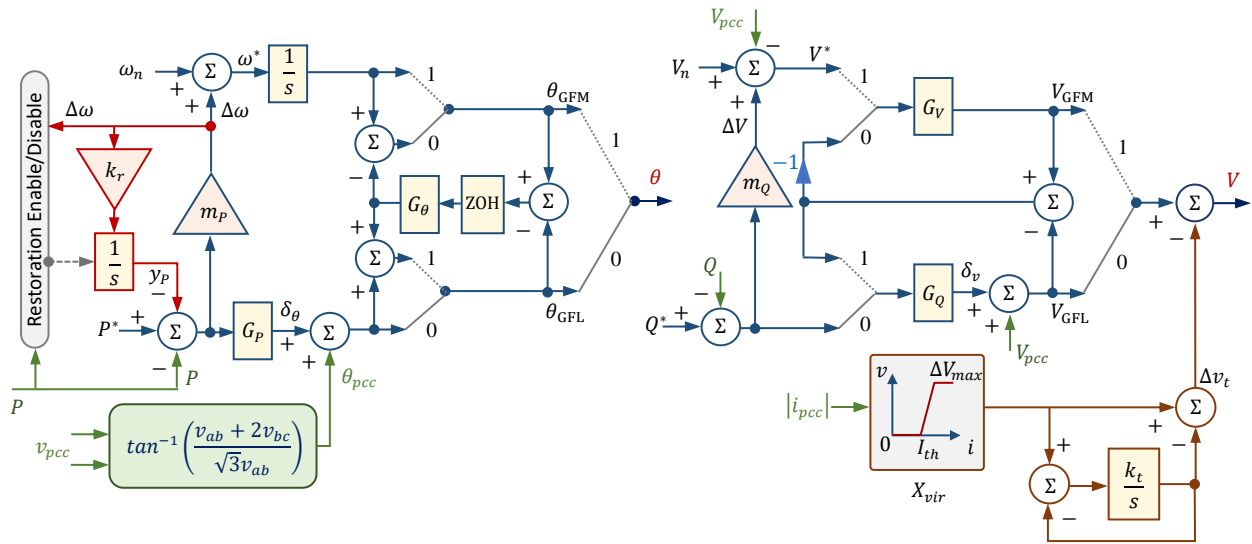

**Fig.5.** Block Diagrams Illustrating the Proposed Control Scheme Applied for Inverters [1]: The left diagram shows the phase-angle reference for inverters. The right diagram displays the voltage-amplitude reference, highlighting the versatility and adaptability of the control strategy in different operational modes.

## 4. Conclusion

In summary, the *Synced Parallel Control Paths* method marks a significant advancement in control systems, offering dual control paths that operate in parallel for enhanced stability and smooth transitions. This method, vital for handling dynamic scenarios, relies on state-space models for tailored control strategies, ensuring effective performance under various conditions. Its practical applications demonstrate seamless transition management without complex protocols, reducing disruptions common in traditional systems. This approach simplifies integration, making it an attractive option in modern control design. This technology represents a significant shift towards more efficient and reliable control in various industries.